\documentclass[aps,amsmath,nofootinbib,amssymb,superscriptaddress,twocolumn]{revtex4}
\usepackage{amssymb,graphics,graphicx,color,epstopdf,subfigure,dcolumn,bm,slashed}
\usepackage[Q=yes,pverb-linebreak=no]{examplep}

\begin{document}

\title{Probing Higgs Boson $CP$ Properties with $t\bar{t}H$
\\
at the LHC and the 100 TeV $pp$ Collider}

\author{Xiao-Gang He} \email{hexg@phys.ntu.edu.tw}
\affiliation{INPAC, SKLPPC, and Department of Physics,
Shanghai Jiao Tong University, Shanghai 200240, China}

\affiliation{Physics Division, National Center for
Theoretical Sciences,  Hsinchu 300, Taiwan}

\affiliation{CTS, CASTS, and
Department of Physics, National Taiwan University, Taipei 10617, Taiwan}

\author{Guan-Nan Li} \email{lgn198741@126.com}
\affiliation{CTS, CASTS, and
Department of Physics, National Taiwan University, Taipei 10617, Taiwan}

\author{Ya-Juan Zheng}\email{yjzheng218@gmail.com}
\affiliation{CTS, CASTS, and
Department of Physics, National Taiwan University, Taipei 10617, Taiwan}

\date{ \today}
\begin{abstract}
The Higgs boson $H$ has the largest coupling to the top quark $t$ among the standard model (SM) fermions. This is one of the ideal places to investigate new physics beyond SM. In this work, we study the potential of determining Higgs boson $CP$ properties at the LHC and future 33 TeV and 100 TeV $pp$ colliders by analysing various operators formed from final states variables in $t\bar{t}H$ production. The discrimination power from SM coupling is obtained with Higgs boson reconstructed from $ H\to \gamma \gamma$ and $ H\to b \bar{b}$. We find that $t\bar{t}b\bar{b}$ process can provide more than 3$\sigma$ discrimination power with 300 $fb^{-1}$ integrated luminosity in a wide range of allowed Higgs to top couplings for the LHC, the 33 TeV and 100 TeV colliders.   For $t\bar{t}\gamma\gamma$ the discrimination power will be below 3$\sigma$ at the LHC, while for 33 TeV and 100 TeV colliders,  more than 3$\sigma$ sensitivity can be reached. 
\end{abstract}

\maketitle


\section{Introduction}

Since the discovery of the Higgs boson at the LHC by the ATLAS and CMS Collaborations~\cite{Chatrchyan:2012ufa,Aad:2012tfa}, experimental measurements have been focusing on the determination of its consistency with standard model (SM) expectation. Within the current experimental and theoretical uncertainties, the discovered Higgs boson is in agreement with the predictions for a SM Higgs boson considering its spin zero property~\cite{Aad:2013xqa, Khachatryan:2014kca}. 
The measurement of Higgs coupling to fermions through $b\bar{b}$~\cite{atlas_bb,cms_hbb}, $\tau^+\tau^-$~\cite{atlas_tautau,cms_tautau}, $t\bar{t}H$~\cite{Aad:2014lma,cms_hbb} 
and to gauge bosons through $WW$~\cite{Chatrchyan:2013iaa,atlas_ww}, $ZZ$~\cite{atlas_zz,Chatrchyan:2013mxa} have also been performed with current LHC data and will be measured more precisely in the future run. The measurement of Higgs self-coupling is also necessary to reconstruct the scalar potential of the Higgs doublet field, while due to the smallness of the signal and large QCD backgrounds the probing of the Higgs self-coupling has to await for a high-luminosity LHC (HL-LHC), 33 TeV, or 100 TeV $pp$ collider~\cite{Dolan:2012rv,Baglio:2012np}.  

It is important to stress that so far all experimental determinations of the Higgs $CP$ properties have been obtained from the Higgs to vector boson couplings with lepton final states. The hypothesis that Higgs boson being a pure pseudoscalar state has been excluded by the present data since the $2\ell2\nu$ and $4\ell$ signals from $H\to WW$ ~\cite{Chatrchyan:2013iaa,atlas_ww} and $ZZ$ ~\cite{atlas_zz,Chatrchyan:2013mxa} decays have been observed. In particular, the angular distributions of the lepton pairs in the $H\to ZZ$ channel are sensitive to the spin-parity of the Higgs boson. Although the present LHC data strongly prefer that $H$ is a $J^{P}=0^+$ state, it is not yet excluded that the Higgs boson has a pseudoscalar component~\cite{Aad:2013xqa,Chatrchyan:2012jja}. Using the present measured signal strengths in various Higgs search channels, theoretical studies on the Higgs $CP$ properties have been performed, and constraints on the mixing angle of the $CP$-even and $CP$-odd component are given model dependently~\cite{Freitas:2012kw,Bhattacharyya:2012tj,Cheung:2013kla,Shu:2013uua,Inoue:2014nva,Bolognesi:2012mm, Englert:2012xt, Djouadi:2013qya,Ellis:2013yxa, Kobakhidze:2014gqa}. 
There are also investigations on how and to what extent to pin down the Higgs $CP$ mixing angle via other Higgs decay channels at the LHC and future linear collider~\cite{Gunion:1996vv,Bhupal Dev:2007is,Berge:2008wi,He:2013tia}.

Among all the Higgs production channels, the investigation of the $t\bar{t}H$ process plays a complementary role in Higgs characterization in the sense that it is very sensitive to the relative magnitudes of the $CP$-even and $CP$-odd top-Higgs Yukawa coupling coefficients.
Although the $t\bar{t}H$ production cross section is only around 1/200 of the inclusive Higgs production for $m_H^{}$ with a mass of $125$ GeV at run I LHC and suffers from large multiplicity of the objects in the final state from top pair decays, the HL-LHC and proposed future hadron colliders with higher collision energy, the high energy LHC (HE-LHC), SppC, etc., are capable of enhancing the $t\bar{t}H$ event rates. In this paper, we will revisit certain operators defined from the momenta of the $t\bar{t}H$ process, which are sensitive to the coefficients of different $CP$-mixed states. We choose two particular Higgs decay modes $\gamma\gamma$ and $b\bar{b}$ for clearness of signal and the wealth of event rates respectively, and study the behaviour of weighted moments at the LHC and the future hadron colliders. We find that these operators for the corresponding $t\bar{t}\gamma\gamma$ and $t\bar{t}b\bar{b}$ processes have a good discrimination power for a $CP$-even Higgs from a $CP$-mixed state.

\section{ effective couplings for the $CP$-mixed Higgs state}\label{sec:tth}

To accommodate possible deviations of Higgs to top quark $t\bar{t}H$ coupling in both strength and $CP$ properties, we parametrize the $t\bar{t}H$  coupling in the following form
\begin{eqnarray}\label{eq:hff}
{\cal L}_{Ht\bar{t}}^{}=-a \frac{m_t}{v} \bar{t}(\cos\xi +i\gamma_5\sin\xi)tH, \label{htt}
\end{eqnarray}
where $v$ is the vacuum expectation value (vev) of Higgs field, and $\sin\xi =0$ and $\cos\xi = 0$ corresponds to the $CP$-even and $CP$-odd coupling coefficient respectively. In particular, the SM Higgs boson has $a=1$ and $\cos\xi =1$.  The angle $\xi$ is an indication of the degree of mixture $CP$ in the physical Higgs $H$. This form of Higgs fermion interaction will be assumed also for other fermions in our discussions.

The dominant Higgs couplings to vector gauge bosons, $W$ and $Z$ will come from the  $CP$-even component of $H$ in the following form\begin{eqnarray}
{\cal L}_{HVV}^{}= a\cos\xi \left (\frac{2m_W^{2}}{v} H W^{\mu}W_\mu + \frac{2m_Z^{2}}{v}H Z^{\mu}Z_\mu \right).
\end{eqnarray}

The main production channel of Higgs boson at the LHC is via gluon fusion process. The effective vertices for a $CP$-mixed Higgs state interacting with gluons at one loop order are given by~\cite{Pilaftsis:1999qt,He:2011ws,Lee:2003nta, Gunion:1989we, Spira:1995rr}
\begin{eqnarray}
{\cal L}_{Hgg}^{}=\left[I_a^gG_{\mu\nu}G^{\mu\nu}+I_b^g\tilde{G}_{\mu\nu}G^{\mu\nu}\right]H,
\end{eqnarray}
with $
I_a^g=a\cos\xi \sum_{i=b,t}F_a(\tau_i),~ I_b^g=a\sin\xi \sum_{i=b,t} F_b(\tau_i)
$ defined as the scalar and pseudoscalar form factors for $Hgg$ which retain the dominant contributions from top and bottom quarks, ${\rm and}~ \tau_i=m_{H}^2/4m_i^2$. $G_{\mu\nu}$ is the gluon field strength and 
$\tilde G_{\mu\nu}$ is the dual of $G_{\mu\nu}$ given by $(i/2)\epsilon_{\mu\nu\alpha\beta} G^{\alpha\beta}$. The two form factors $F_a(\tau)$ and $F_b(\tau)$ can be expressed in terms of scaling function $f(\tau)$ as
\begin{eqnarray}
F_a(\tau)=\tau^{-1}\left[1+(1-\tau^{-1})f(\tau)\right],\quad
F_b(\tau)=\tau^{-1}f(\tau),
\end{eqnarray}
where
\begin{eqnarray}
f(\tau)=
\begin{cases}
\left[\sin^{-1}\left(\sqrt{\tau}\right)\right]^2,& \text{if}\quad \tau\leq 1,\\
-\frac{1}{4}\left[\ln(\frac{\sqrt{\tau}+\sqrt{\tau-1}}{\sqrt{\tau}-\sqrt{\tau-1}})-i\pi\right]^2,& \text{if}\quad\tau>1.
\end{cases}
\end{eqnarray}

The Higgs to diphoton effective coupling is formulated as~\cite{He:2011ws,Lee:2003nta, Gunion:1989we, Spira:1995rr}
\begin{eqnarray}
{\cal L}_{H\gamma\gamma}^{}=\left[I_a^\gamma F_{\mu\nu}F^{\mu\nu}+I_b^\gamma\tilde{F}_{\mu\nu}F^{\mu\nu}\right]H,
\end{eqnarray}
where $F_{\mu\nu}$ and $\tilde F_{\mu\nu}$ are the photon field strength and its dual, respectively. The 
form factors  $I^\gamma_{a,b}$ are given by $I_a^{\gamma}=a\cos\xi [2\sum_{i=b,t}N_C^{}Q_i^2F_a(\tau_i)-F_1(\tau_W^{})]$, ~$I_b^\gamma=2a\sin\xi \sum_{i=b,t} N_C^{}Q_i^2F_b(\tau_i)
$ with $N_C^{}=3$ the colour factor for quarks and $Q_i$ the electric charge of the corresponding particle.
$F_1(\tau_W^{})$ is from $W$ boson loop contribution with
 \begin{eqnarray}
 F_1(\tau)=2+3\tau^{-1}+3\tau^{-1}(2-\tau^{-1})f(\tau).
 \end{eqnarray}

\section{Production of $t\bar{t}H$ at $pp$ Colliders}

Deviations of Higgs boson coupling to top quark from SM one can show up in different ways. We study how to identify such deviations using $pp \to  t \bar t HX$ with the Higgs boson identified by $H\to b\bar{b}$ and $H\to\gamma\gamma$ since the former has the largest branching ratio and  the latter has clearer signals compared with other decay modes.

Both ATLAS~\cite{Aad:2014lma,atlas_hbb} and CMS~\cite{cms_hbb} have performed experimental searches on the $t\bar{t}H$ production process, with $H$ decays into $\gamma\gamma$ and $b\bar{b}$ separately. For the $t\bar{t}b\bar{b}$ process, collecting the $\sqrt{s}=8$ TeV data final state are categorized according to their jet and $b$-tagged jet multiplicities.
The observed cross section upper bound is $4.1\sigma_{\rm SM}$~\cite{atlas_hbb}, and $3.3\sigma_{\rm SM}$~\cite{cms_hbb}. For the $t\bar{t}\gamma\gamma$ process, the observed upper limit on cross section is $6.7\sigma_{\rm SM}$~\cite{Aad:2014lma}. We still need to wait for more accurate data to draw conclusions about the production cross section. 

\begin{figure}[t]
\begin{centering}
\begin{tabular}{c}
\includegraphics[width=0.45\textwidth]{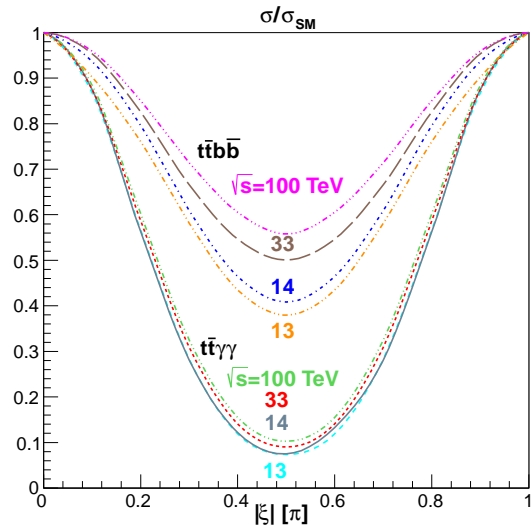}
\end{tabular}
\caption{The cross section ratio $\sigma/\sigma_{\rm SM}$ versus $CP$ mixing angles for $t\bar{t}\gamma\gamma$ and $t\bar{t}b\bar{b}$ processes at $\sqrt{s}$= 13 TeV, 14 TeV, 33 TeV, and 100 TeV. In the figure, the parameter $a$ is set to be 1.}\label{fig:strength}
\end{centering}
\end{figure}

The interaction in Eq.(\ref{htt}) can cause deviation in $t\bar{t}H$ cross section from SM prediction and in the Higgs decays. We have evaluated the $t\bar{t}H$ cross sections using {MadGraph5}\Q{_}{aMC} particularly {\tt heft} model where the Higgs effective theory is included~\cite{Alwall:2014hca} with a mild rapidity cuts on the top and antitop quarks $|\eta_{t,\bar{t}}|<4$. The results with the parameter $a=1$ and range of $\xi$ from 0 to $\pi$ and collision energy $\sqrt{s}=$ 13 TeV, 14 TeV, 33 TeV and 100 TeV with $m_H^{}=125$ GeV are shown in Fig.~\ref{fig:strength}. Here we would like to point out that  one should consistently take the high-order correction to the cross section into account. For $t\bar{t}H$ process, the NLO prediction for a Higgs boson mass of 125 GeV has cross section around 86 fb, 130 fb, 611 fb and 33700 fb~\cite{Dittmaier:2011ti,Beenakker:2001rj,Dawson:2002tg,Frixione:2014qaa,Demartin:2014fia} at $\sqrt{s}=7$ TeV, 8 TeV, 14 TeV and 100 TeV.  Estimated using CTEQ6.6, the NLO corrections are positive. This is higher than that obtained with the cross section of about 72 fb, 107 fb, 476 fb and 24010 fb using the LO results. A $K$-factor of 1.19 - 1.40 ~\cite{Dittmaier:2011ti,Frixione:2014qaa,Demartin:2014fia} should be applied when we use the values shown in Fig.~\ref{fig:strength}.

With $a$ fixed at 1, one obtains the cross sections for the SM for $\xi =0$. When going away from $\xi =0$, one can clearly see that the cross sections become considerablely smaller than those of the SM predictions. One wonders this fact can be used to distinguish the SM $t\bar{t}H$ coupling with that from a beyond SM. 
Global fit on the $CP$-violating phase using the observed rates of $gg\to H$ and $H\to\gamma\gamma$ with the LHC 7 and 8 TeV data can be translated into constraints on $\xi$ as $|\xi|<0.75\pi$ at $99.73\%$ C.L assuming $a=1$~\cite{Djouadi:2013qya}.
 It has been shown that if the LHC measurement of $\sigma(t\bar{t}H)$ is with an accuracy of $20\%$, it is indicated that $|\xi|<0.17\pi$ can be obtained~\cite{Ellis:2013yxa}. When the cross section is measured to good precision, one can obtain some important information about the interaction.

If experiments obtain a cross section different from that with $a=1$ and $\xi=0$, one immdediately knows that beyond SM physics is needed.  However, should a cross section close to the SM will be obtained, one cannot rule out the possibility of beyond the SM $t\bar{t}H$ coupling given in Eq.(\ref{htt}). This is because that even with a $\xi$ nonzero and therefore  smaller cross sections as shown in Fig.~\ref{fig:strength} if the parameter can be varied from 1, one can still adjust the value of $a^2$ to be inverse of $\sigma/\sigma_{\rm SM}$ to obtain similar values as SM ones. One, however, note that just measuring cross sections, one will not be able to have information about Higgs $CP$ properties. Therefore it is desirable to find ways to distinguish SM from beyond SM $t\bar t H$ couplings independent of the overall scaling parameter $a$ and also to provide information about Higgs $CP$ properties. In the rest of this paper, we show that certain weighted moments in $pp \to  t\bar{t}HX$ process are sensitive to the relative magnitudes of the $CP$-even and $CP$-odd interactions in $ t\bar t H$. Defined from the ratios of integrated operators to total cross sections, these weighted moments are also partially free from NLO corrections, where the $K$ factors can be factored out.

\section{ Weighted moments in $ t\bar{t}H$ process}\label{sec:operator}

To achieve this, we take the same approach as those proposed in Ref. \cite{Gunion:1996xu} to study operators in the following formed from final product variables which are sensitive to $\cos^2\xi-\sin^2\xi$
\begin{eqnarray}\label{eq:operator}
&&{\cal O}_{1}^{}\equiv \frac{(\vec{p}_t\times \hat{n})\cdot (\vec{p}_{\bar{t}}\times \hat{n})}{|(\vec{p}_t\times \hat{n})\cdot (\vec{p}_{\bar{t}}\times \hat{n})|},\quad{\cal O}_{2}^{}\equiv \frac{p_t^xp_{\bar{t}}^x}{|p_t^xp_{\bar{t}}^x|}\nonumber\\
&&{\cal O}_{3}^{}\equiv  \frac{(\vec{p}_t\times \hat{n})\cdot (\vec{p}_{\bar{t}}\times \hat{n})}{p_t^Tp_{\bar{t}}^T},\quad
{\cal O}_{4}^{}\equiv  \frac{(\vec{p}_t\times \hat{n})\cdot (\vec{p}_{\bar{t}}\times \hat{n})}{|\vec{p}_t||\vec{p}_{\bar{t}}|},\nonumber\\
&&{\cal O}_{5}^{}\equiv \frac{p_t^xp_{\bar{t}}^x}{p_t^Tp_{\bar{t}}^T},\quad
{\cal O}_{6}^{}\equiv \frac{p_t^zp_{\bar{t}}^z}{|\vec{p}_t||\vec{p}_{\bar{t}}|},
\end{eqnarray}
where $p_{t,\bar{t}}^T$ denote the magnitudes of the $t$ and $\bar{t}$ transverse momenta, $\hat{n}$ is a unit vector in the direction of the beam line and defines the $z$ axis, while $x$ axis is chosen to be any fixed direction perpendicular to the beam. There may be other operators but we will restrict ourselves to the six operators in Eq.~(\ref{eq:operator}) as examples.

\begin{figure}[t]
\begin{centering}
\begin{tabular}{c}
\includegraphics[width=0.45\textwidth]{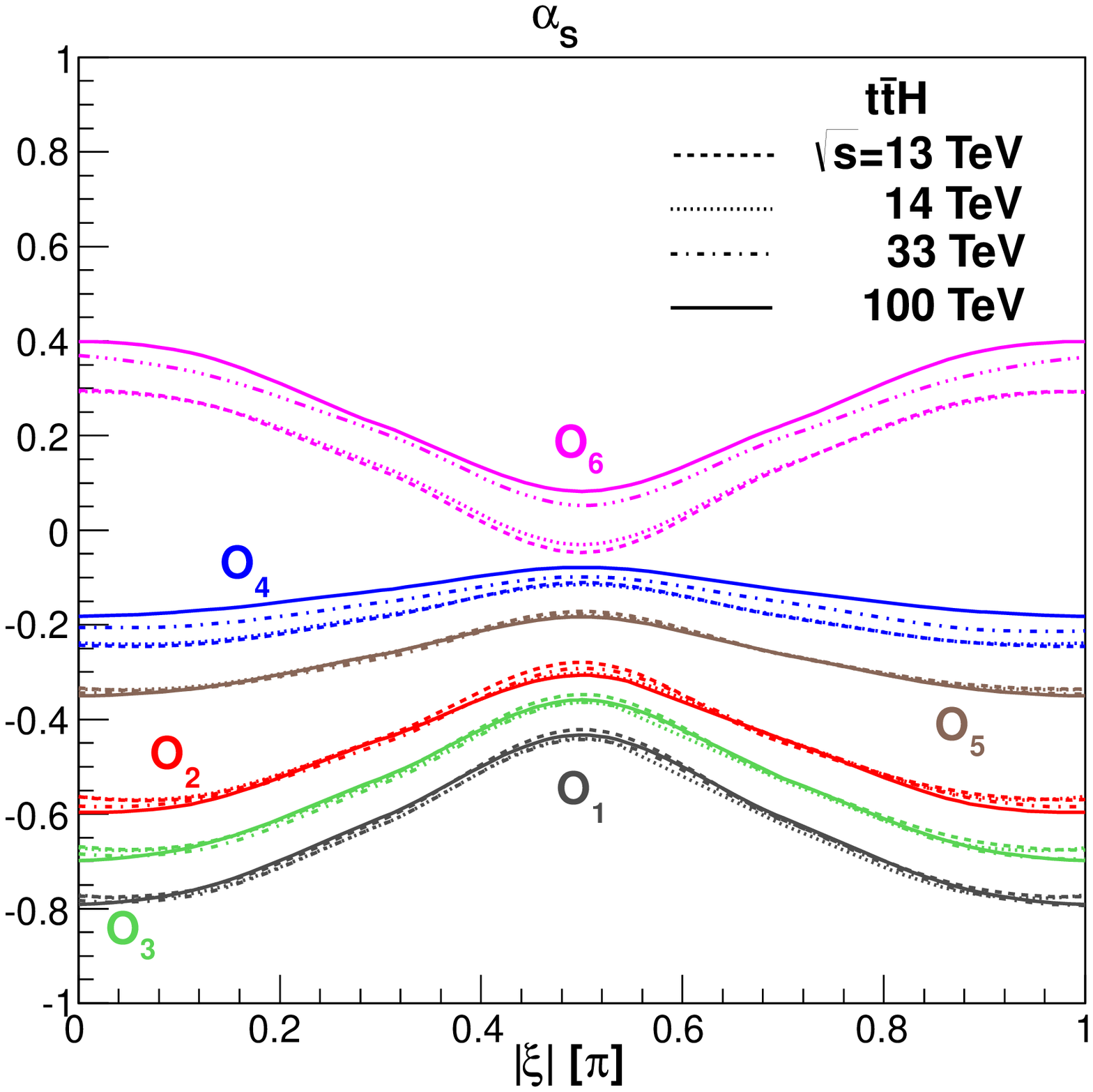}\\
\includegraphics[width=0.45\textwidth]{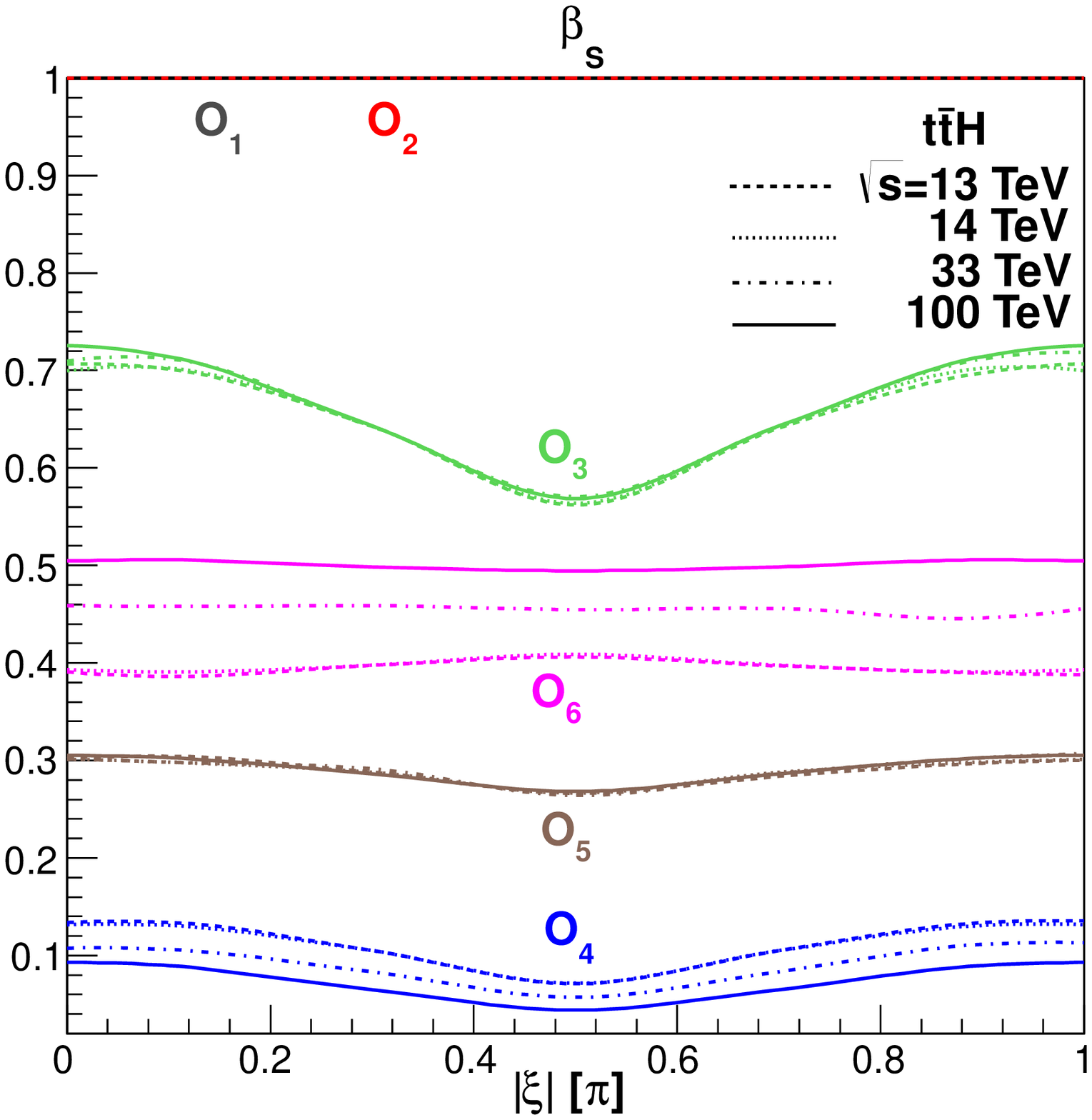}
\end{tabular}
\caption{The $\alpha_S$ and $\beta_S$ as functions of $\xi$ for $pp \to  t\bar t HX$ process at $\sqrt{s}$= 13 TeV, 14 TeV, 33 TeV and 100 TeV. }\label{fig:alpha}
\end{centering}
\end{figure}

As mentioned before that the unknown parameter $a$ may make interpretation of information extracted difficult, it is therefore desirable to find observables that are independent of the parameter $a$. To this end we define the following observables by taking ratios to remove the $a$ dependence,
\begin{eqnarray}\label{eq:alpha}
\alpha[{\cal O}_i^{}]\equiv \frac{\int [{\cal O}_i^{}]\{ d\sigma (pp\to t\bar{t}XX)/d{R}\}d{ R}}{\int \{d\sigma (pp\to t\bar{t}XX)/d{ R}\}d{R}},
\end{eqnarray}
where $XX$ represents Higgs decay products and $R$ is the phase space of $t\bar{t}XX$ process. The values of $\alpha_S^{}$ are shown in the upper panel of Fig.~\ref{fig:alpha}. One can see that the changes of $\alpha_S$ against $\xi$ is significant providing hope to distinguish Higgs boson coupling with top quark with mixed $CP$ component.

\section{Discrimination power for  $t\bar{t} \gamma \gamma$ and $t\bar{t} b \bar{b}$ process}\label{sec:D}

We now study to what accuracy for a given $\alpha_S^{}$ it can be measured.
One must consider errors associated with the variable when both signal and SM background are taken into account. The statistic error for $\alpha_S^{}$ is given by
\begin{eqnarray}\label{eq:del_alpha}
\delta\alpha_S^{}=\frac{1}{\sqrt{S}}\left[\beta_S-\alpha_S^2+\frac{B}{S}(\beta_B-2\alpha_B\alpha_S+\alpha_S^2)\right]^{1/2},
\end{eqnarray}
with $S$ and $B$ the total number of events for signal and background process respectively.
The variable $\beta_{S,B}$ for a given operator $O_i$ is defined as
\begin{eqnarray}
\beta[{\cal O}_i^{}]\equiv \frac{\int [{\cal O}_i^{}]^2\{ d\sigma (pp\to t\bar{t}XX)/dR\}dR}{\int\{ d\sigma (pp\to t\bar{t}XX)/dR\}dR},
\end{eqnarray}
and shown in the lower panel of Fig.\ref{fig:alpha} for different operators varing the $CP$ mixing angle $\xi$ at different collision energies. Note that $\beta_S^{}({\cal O}_{1,2}$) are equal to 1.

To quantify the ability to distinguish the SM pure scalar case from any $CP$-mixed Higgs state of each operator, the discrimination power $D$ is defined
\begin{eqnarray}\label{eq:D}
D\equiv\frac{|\alpha_S^{\rm SM} (\xi=0)-\alpha_S^{\xi}(\xi)|}{\delta\alpha_S^{\rm SM}(\xi=0)}.
\end{eqnarray}
%
We vary the mixing angle $\xi$ from 0 to $\pi$ when a series of discrimination powers can be obtained accordingly from Eq.(\ref{eq:D}).  

For a full consideration, we must include both signal and background for Higgs and top pair decay.  
Since $H\to b\bar{b}$ has the largest branching ratio and $H\to\gamma\gamma$ has clearer signal, we will study the behaviours of $\alpha_S^{}$ for the $t\bar{t}b\bar{b}$ and $t\bar{t}\gamma\gamma$ processes aiming for the potential of these operators to discriminate a nonzero $\xi$ compared with SM coupling in light of new data available at different energies for a $pp$ collider. 

We compute the cross section of the $t\bar{t}\gamma\gamma$ and $t\bar{t}b\bar{b}$ processes  using {MadGraph5}\Q{_}{aMC}~for both signal and background \cite{Alwall:2014hca}. Cross sections for $t\bar{t}\gamma\gamma$ production are obtained after basic cuts $|\eta_{t,\bar{t},\gamma}|<4$, $p_\gamma^T>25$ GeV and the mass window cut $|M_{\gamma\gamma}-m_H^{}|<5$~GeV, with $m_H^{}=125$ GeV.  For the $ttb\bar{b}$ process, we impose the basic cuts with $|\eta_{t,\bar{t},b,\bar{b}}|<4$ and $p_{b,\bar{b}}^T>25$ GeV. 
%
%
\begin{figure*}[t]
\begin{centering}
\begin{tabular}{c}
\includegraphics[width=0.45\textwidth]{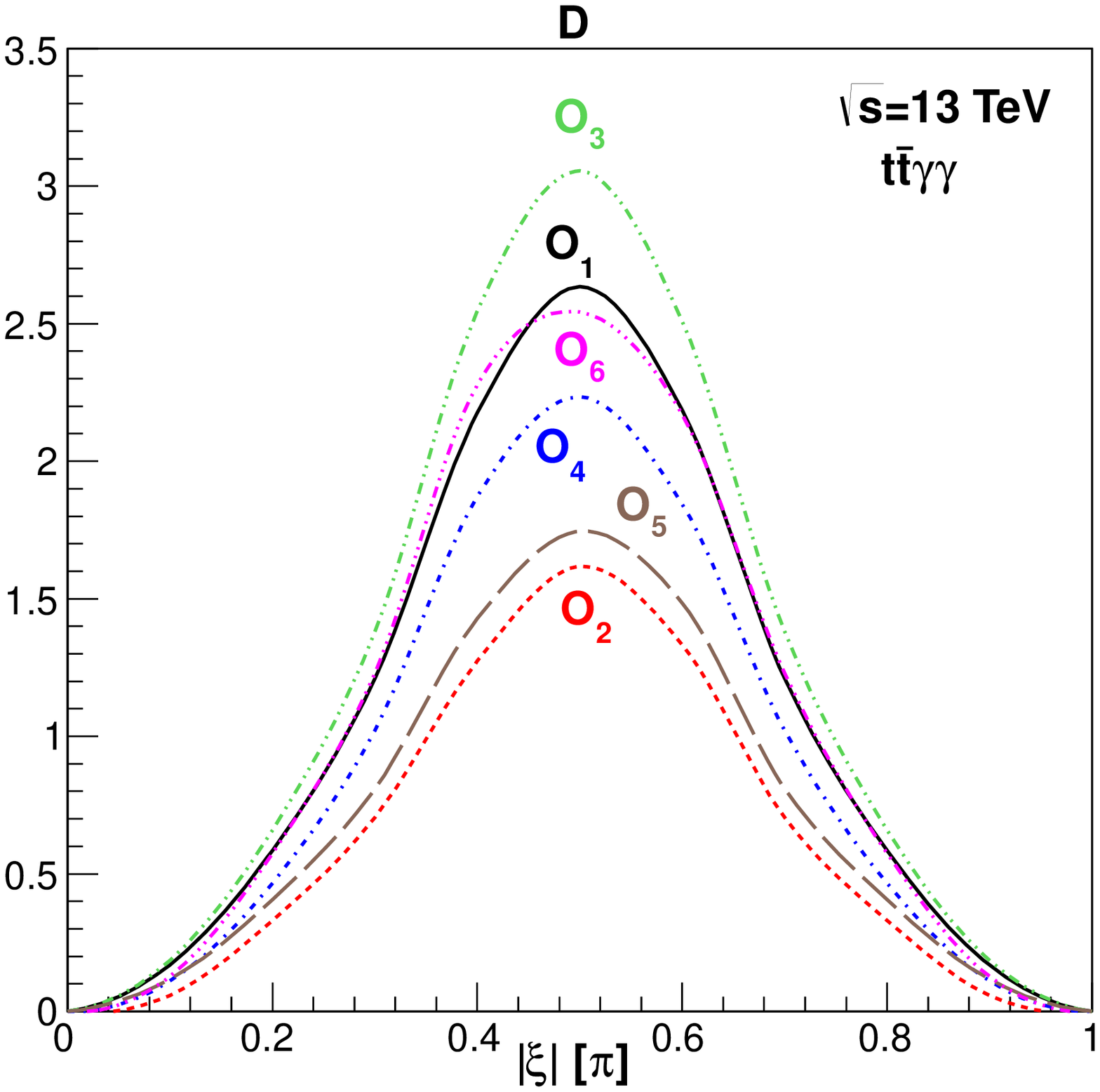}
\includegraphics[width=0.45\textwidth]{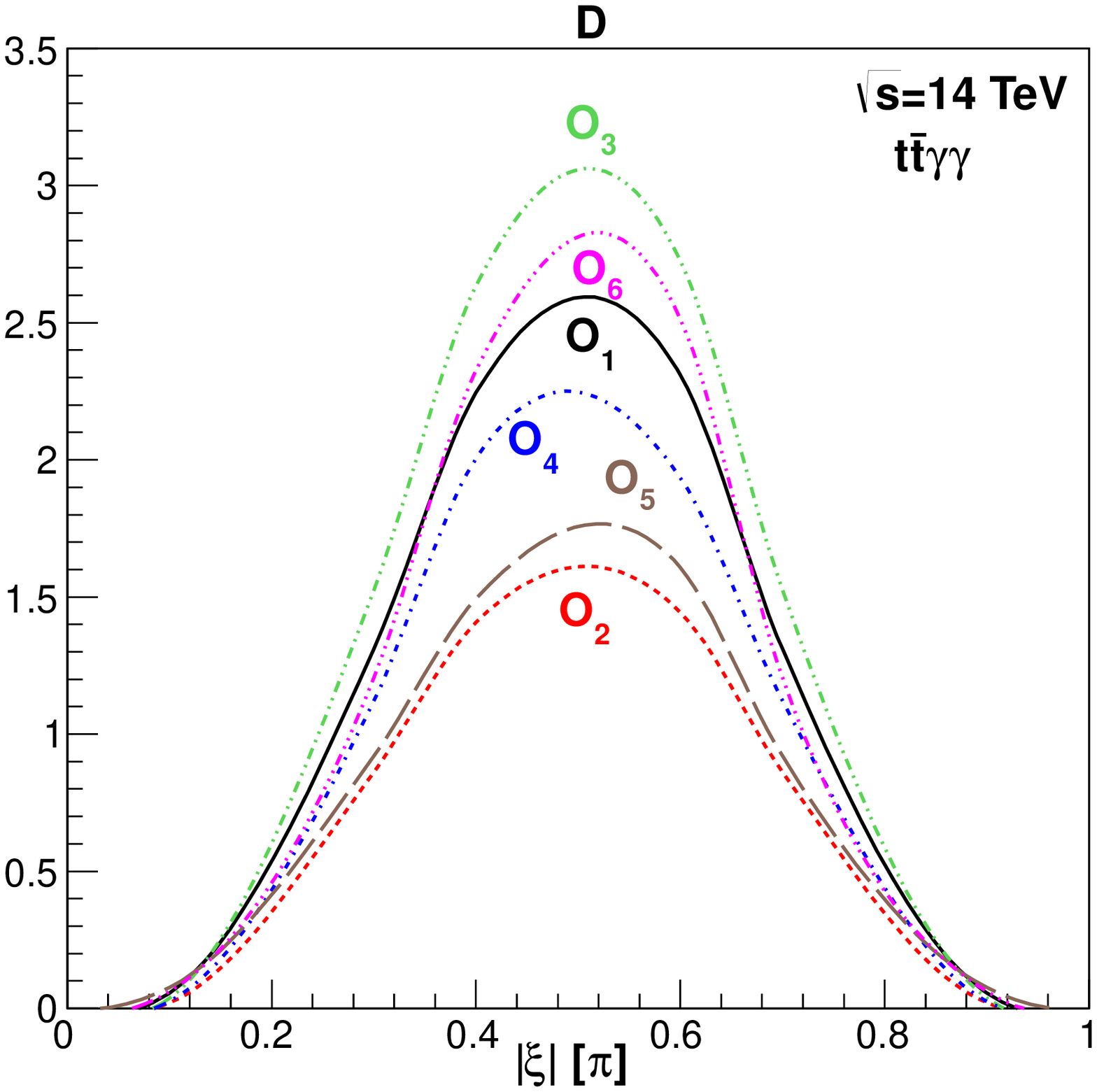}\\
\includegraphics[width=0.45\textwidth]{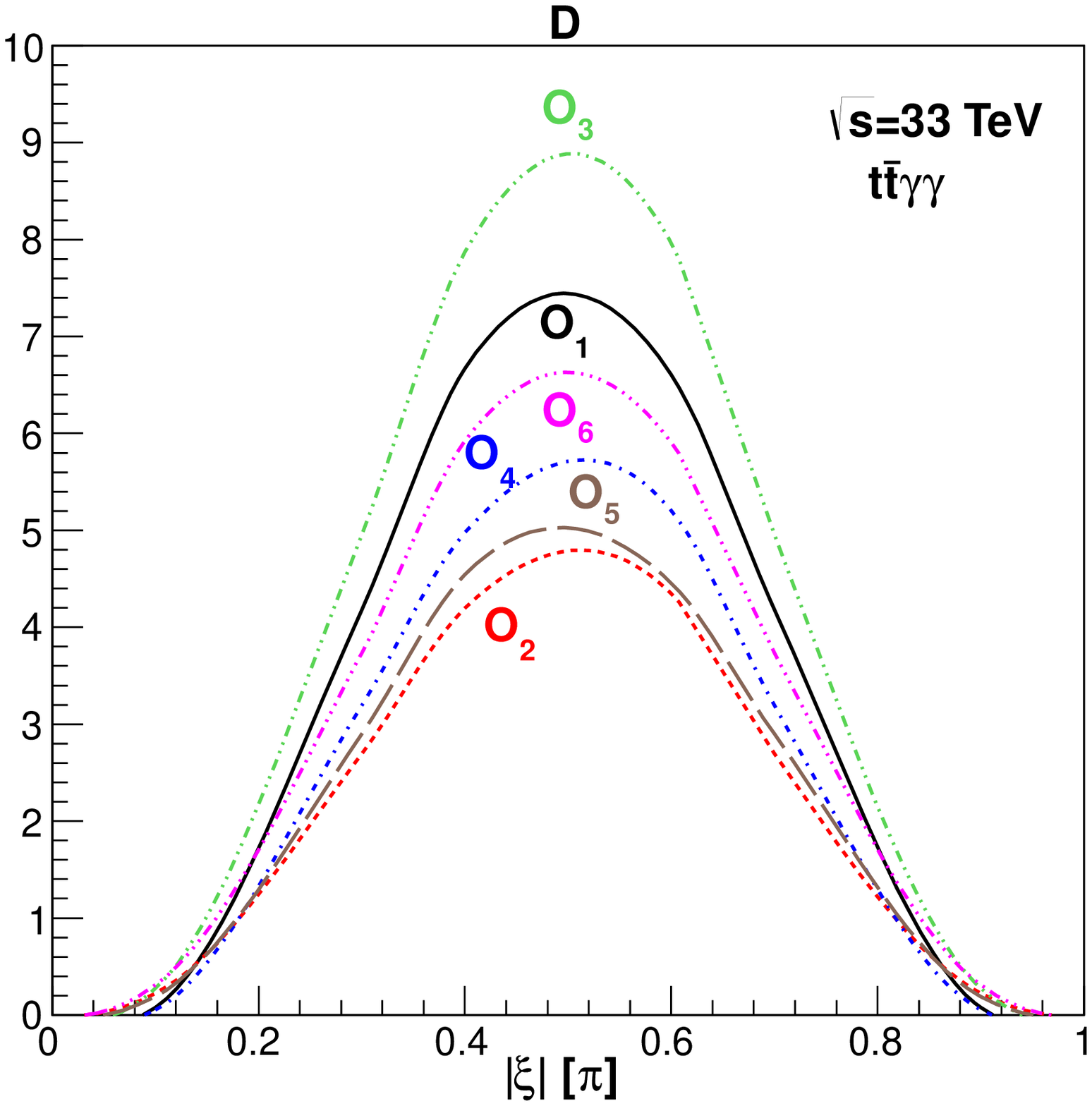} 
\includegraphics[width=0.45\textwidth]{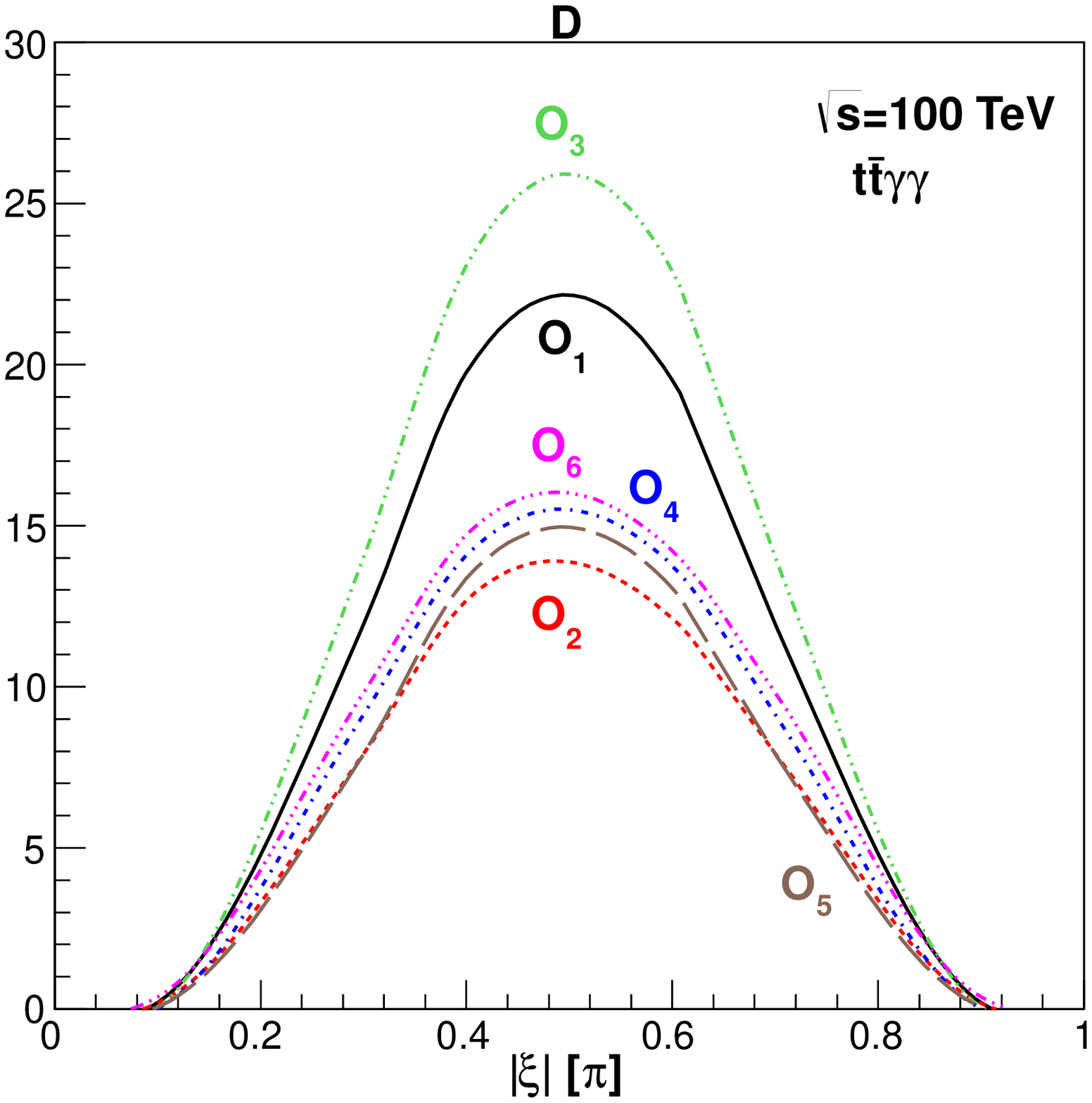}
\end{tabular}
\caption{The discrimination power $D$ versus $CP$ mixing angles $\xi$ for each operator for $t\bar{t}\gamma\gamma$ process at $\sqrt{s}$=13 TeV, 14 TeV, 33 TeV and 100 TeV . 
}\label{fig:D_ttrr}
\end{centering}
\end{figure*}
\begin{figure*}[t]
\begin{centering}
\begin{tabular}{c}
\includegraphics[width=0.45\textwidth]{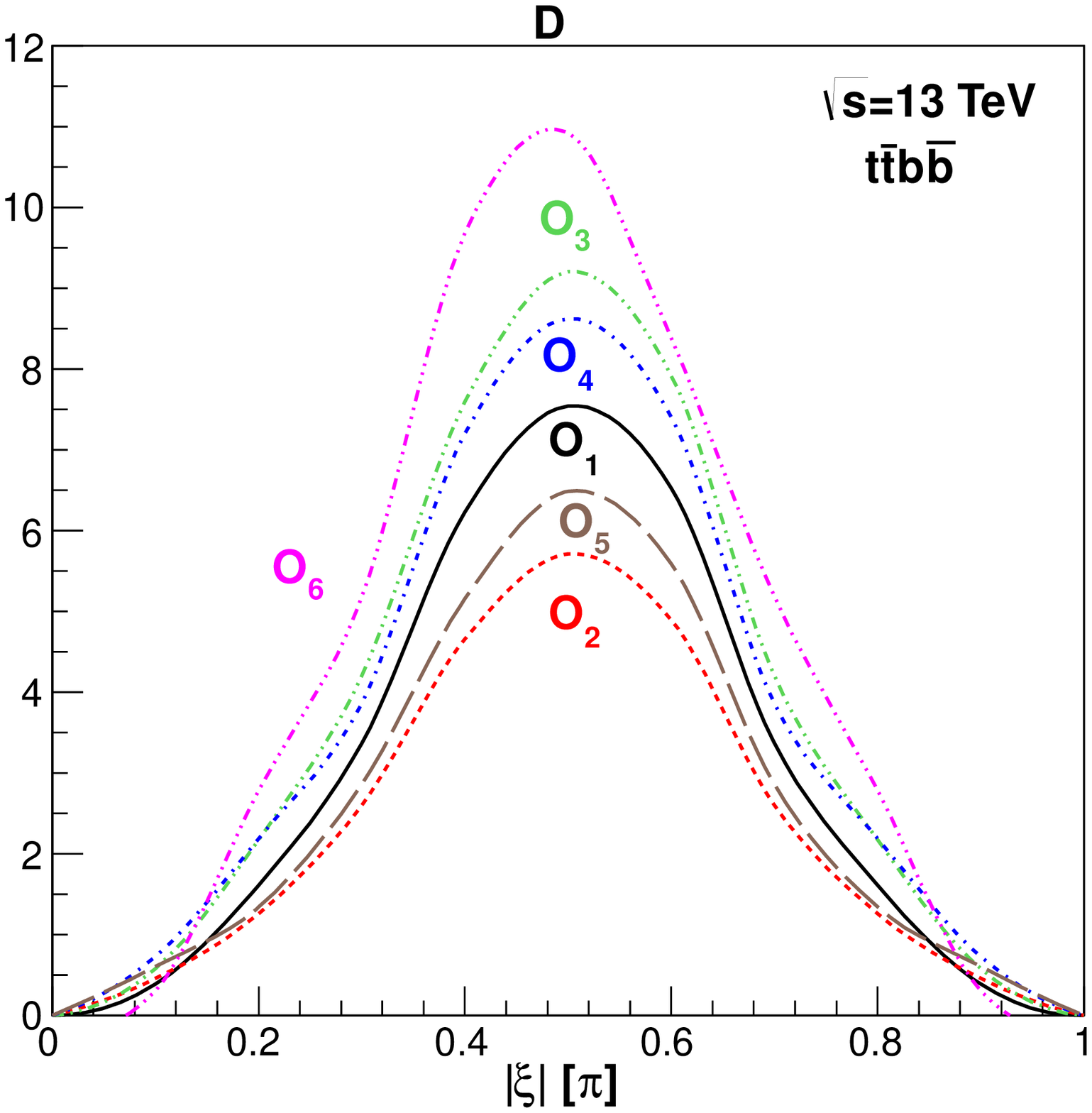}
\includegraphics[width=0.45\textwidth]{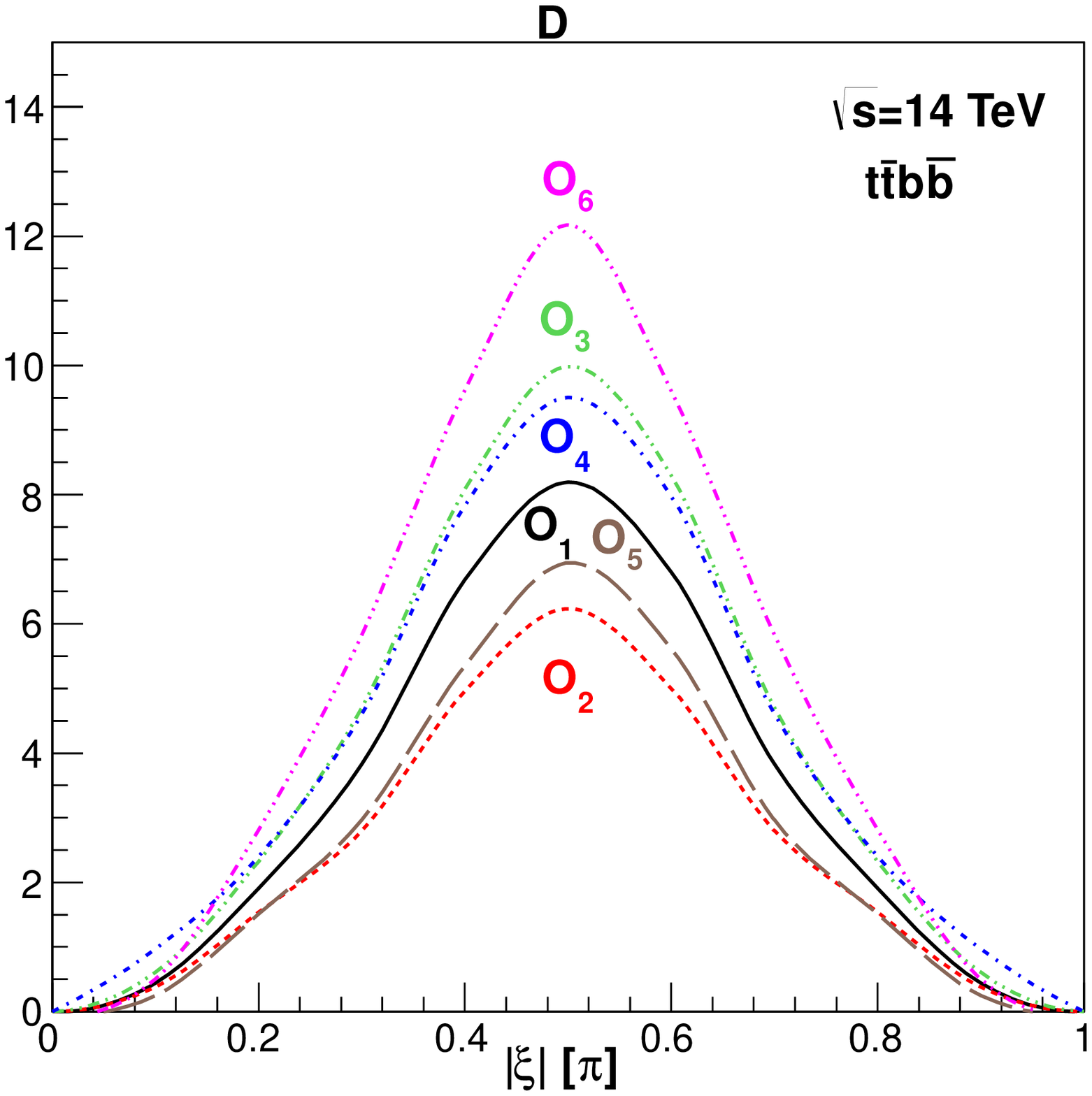}\\
\includegraphics[width=0.45\textwidth]{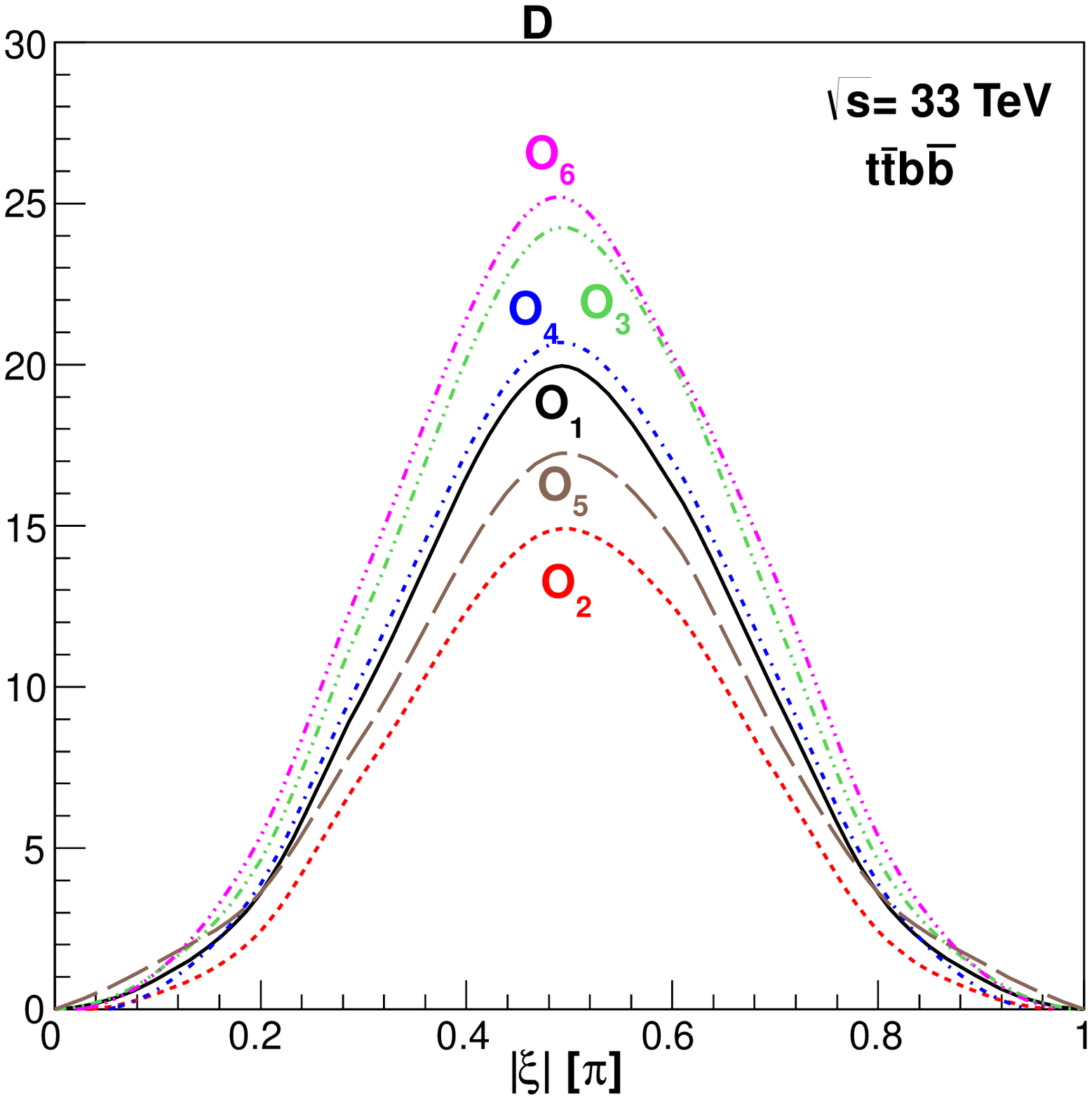} 
\includegraphics[width=0.45\textwidth]{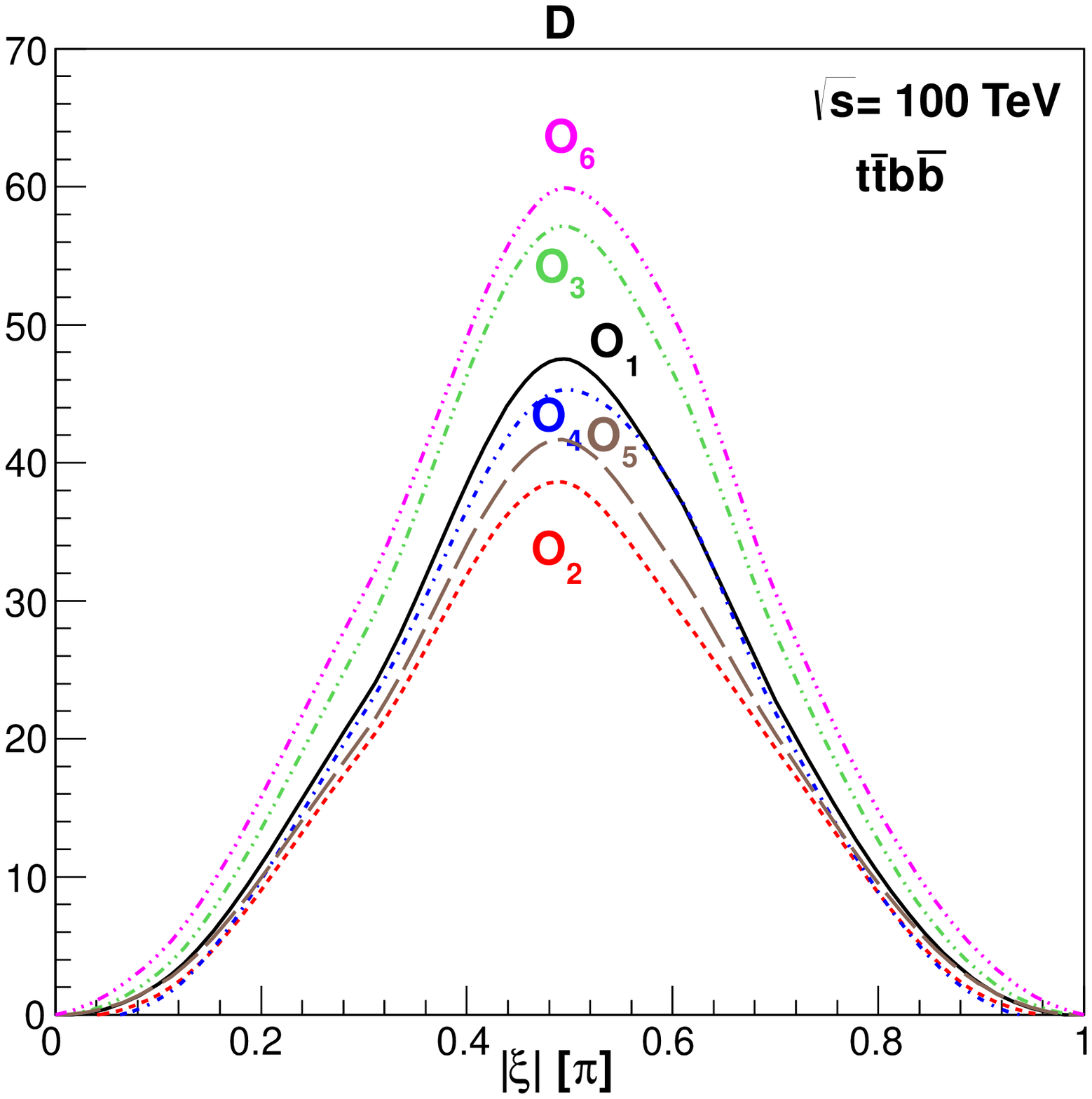}
\end{tabular}
\caption{The discrimination power $D$ versus $CP$ mixing angles $\xi$ for each operator for $t\bar{t}b\bar{b}$ process at $\sqrt{s}=13$ TeV, 14 TeV, 33 TeV and 100 TeV. 
}\label{fig:D_ttbb}
\end{centering}
\end{figure*}

If we further include the $t\bar{t}$ pair decay, a total factor of the change on the cross section $\sigma$ and discrimination power $D$ can be estimated by the corresponding decay branching ratios. We choose top quark decays semileptonically and antitop quark decays hadronically in order for the identification of $t$ and $\bar{t}$ and also the reconstruction of the transverse momentum of $t$ and $\bar{t}$. Therefore, the decay products from top pair will be $2b$-jet $+2j+\ell+\slashed E_T$, the reduced signal cross section can be estimated by
$
n\equiv {\cal B}(t\to b W)^2{\cal B}(W\to l \nu){\cal B}(W\to j j)\simeq \left(\frac{9}{10}\right)^2 \times\frac{1}{3}\times\frac{2}{3}
,
$
with $n\simeq 0.18$ for $t\bar{t}\gamma\gamma$ process.
While for $t\bar{t}b\bar{b}$ cross section, an additional suppression factor from $b$-tagging efficiency ($\sim 60\%$) should be included leading to
$n(t\bar{t}b\bar{b})\simeq 0.0648$. Consequently, by including top pair decay the value of $D$ will be affected by a total suppression factor of $\sqrt{n}\simeq$ 0.42 (0.25) for $t\bar{t}\gamma\gamma$ ($t\bar{t}b\bar{b}$) from Eq.(\ref{eq:D}).

Given $\alpha_S^{}$, $\beta_S^{}$, $\alpha_B^{}$, $\beta_B^{}$ and background event number $B$, it is straightforward to compute the required signal event number $S$ for a SM-like Higgs to achieve certain value of $D$. For example, if one wants to discriminate $t\bar{t}H$ coupling of $\xi=0$ from $\xi=0.3\pi$ at the $3\sigma$ statistical level, the prediction from $t\bar{t}\gamma\gamma$ and $t\bar{t}b\bar{b}$ are different. The $t\bar{t}\gamma\gamma$ background event number is $B\sim50$ (2280) at $\sqrt{s}=13$ (100) TeV, estimated from the operator with the largest discrimination power ${\cal O}_3$ we find that $D=3$ requires signal event number to be around $750$ (1950). The simulated value for LO signal rates are about 160 (9870). Therefore, using $t\bar{t}\gamma\gamma$ channel it is difficult to discriminate the $t\bar{t}H$ of $\xi=0$ coupling from a $\xi=0.3\pi$ mixed coupling at $3\sigma$ statistical level at 13 TeV. However, a future $pp$ collider at 100 TeV can produce enough signal events to achieve the discrimination at larger than $3\sigma$ level. The situation is better in $t\bar{t}b\bar{b}$ channel. The SM $t\bar{t}b\bar{b}$ background has a event number of around $8.0\times 10^5$ (7.8$\times10^7$) at $\sqrt{s}=13$ (100) TeV,
 estimated from the operator with largest discrimination power ${\cal O}_6$ we find that $D=3$ requests signal events number to be $4.1\times 10^4$ ($4.3\times10^5$). The real LO signal event number is around $7.1\times10^4$ ($4.4\times10^6$), which is significantly lager than required. Consequently, from the $t\bar{t}b\bar{b}$ process, one is able to discriminate the $t\bar{t}H$ coupling with $\xi=0$ from a $\xi=0.3\pi$ $CP$-mixed state at more than $3\sigma$ statistical level with any collision energy larger than 13 TeV at $pp$ collider.
 
The figures~\ref{fig:D_ttrr} and~\ref{fig:D_ttbb} show the corresponding $D$ values for $t\bar{t}\gamma\gamma$ and $t\bar{t}b\bar{b}$ final state after including the above decay and $b$-tagging efficiencies with the $CP$ mixing angle running from $0$ to $\pi$ with collision energy at 13 TeV, 14 TeV, 33 TeV and 100 TeV. 
For both processes, the number of events are obtained with integrated luminosity assumed to be ${\cal L}= 300~{\rm fb}^{-1}$ and an enhancing factor of $\sqrt{10}\sim3.2$ on $D$ will appear if we consider the HL-LHC with ${\cal L}=$3 ab$^{-1}$. 
Comparison between each two operators shows that ${\cal O}_1,{\cal O}_3,{\cal O}_4$, and ${\cal O}_6$ are most useful balancing the sensitivity and errors, among which ${\cal O}_3$ probably performs the best. ${\cal O}_3$ is also easy to be constructed which is irrelevant to the $z$-component of top and antitop momentum $p_{t,\bar{t}}$. 

The results shown in Fig.~\ref{fig:D_ttrr} tell that for $t\bar{t}\gamma\gamma$ process at $\sqrt{s}$= 13 TeV and 14 TeV  the discrimination powers for operators ${\cal O}_1$-${\cal O}_6$ are generally less than three in the allowed $CP$ mixing region. 
When the integrated luminosity is increased to be 3 ab$^{-1}$ at HL-LHC, $D\gtrsim3$ can be reached in the region $|\xi|\gtrsim0.3\pi$ for all operators. It is much more optimistic at $\sqrt{s}=33$ TeV and $100$ TeV, which shows that all the operators can give large enough statistical significance in the region $|\xi|\gtrsim0.2\pi$. The operator ${\cal O}_3$ gives the best discrimination power irrelevant of the collision energy and $CP$ mixing angle. 
All the operators reach maximal discrimination power at $|\xi|=0.5\pi$.

Comparing the $t\bar{t}b\bar{b}$ process with $t\bar{t}\gamma\gamma$, we want to emphasize, as can be seen from Fig.~\ref{fig:D_ttbb}, that $H\to b\bar{b}$ decay modes shows better discrimination powers due to the large event rate of the Higgs to $b\bar{b}$ decay branching ratio. For $t\bar{t}b\bar{b}$ process, the operator ${\cal O}_3$ and ${\cal O}_6$ give the best discrimination powers.
Note that if we make full analysis considering final state jets and leptons from top quark decay, and impose further cuts on the jet multiplicities, the value of discrimination power $D$ will appear even better than present numbers. 

\section{summary and discussions}\label{sec:sum}

The Higgs boson $H$ has the largest coupling to the top quark $t$. The $t\bar{t}H$ interaction can be sensitive to the investigation of new physics beyond SM.
In this work, we have studied the potential of determining Higgs boson $CP$ properties at the LHC and future 33 TeV and 100 TeV $pp$ coliders by analysing various operators formed from final states variables in $t\bar{t}H$ production. 

A $CP$ violating Higgs coupling to top quark can cause significant deviations from SM predictions for $pp \to  t\bar t HX$. We have evaluated the $t\bar{t}H$ cross sections using {MadGraph5}\Q{_}{aMC} particularly {\tt heft} model. We find that the cross sections become considerably smaller than those of the SM predictions when CP violating component gets bigger. This fact may be used to distinguish the SM $t\bar{t}H$ coupling with that from a beyond SM.  But this may be caused by an overall scaling factor to the coupling. To further identify the relative magnitudes of the $CP$-even and $CP$-odd interactions in $ t\bar t H$,  certain sensitive weighted moments in $pp \to  t\bar{t}HX$ process can help to achieve this.  
We obtained results for discrimination power for several operators with the Higgs boson identified by $H\to b\bar{b}$ and $H\to\gamma\gamma$.
For an integrated luminosity is 300 $fb^{-1}$, we find that, with $t \bar t H \to t\bar{t}\gamma\gamma$, the discrimination power will be below 3$\sigma$ at the LHC, while for 33 TeV and 100 TeV colliders,  more than 3$\sigma$ sensitivity can be reached. On the other hand, $t \bar t H \to t\bar{t}b\bar{b}$ process can provide more than 3$\sigma$ discrimination power in a wide range of allowed Higgs to top couplings for the LHC, the 33 TeV and 100 TeV colliders.

\acknowledgements
  The work was supported in part by MOE Academic Excellent Program (Grant No.
102R891505) and MOST of ROC, and in part by NNSF(Grant No.11175115) and Shanghai
Science and Technology Commission (Grant No.11DZ2260700) of PRC.


\begin{thebibliography}{99}

\bibitem{Chatrchyan:2012ufa}
  S.~Chatrchyan {\it et al.}  [CMS Collaboration],
  Phys.\ Lett.\ B {\bf 716}, 30 (2012)
  [arXiv:1207.7235 [hep-ex]].

\bibitem{Aad:2012tfa}
  G.~Aad {\it et al.}  [ATLAS Collaboration],
  Phys.\ Lett.\ B {\bf 716}, 1 (2012)
  [arXiv:1207.7214 [hep-ex]].

\bibitem{Aad:2013xqa} 
  G.~Aad {\it et al.}  [ATLAS Collaboration],
  Phys.\ Lett.\ B {\bf 726}, 120 (2013)
  [arXiv:1307.1432 [hep-ex]].
  
\bibitem{Khachatryan:2014kca} 
  V.~Khachatryan {\it et al.}  [ CMS Collaboration],
  arXiv:1411.3441 [hep-ex].
  
       \bibitem{atlas_bb}
 [ATLAS Collaboration], ATLAS-CONF-2013-079.
 
     \bibitem{cms_hbb}
 [CMS Collaboration], CMS PAS HIG-14-010.
 
        \bibitem{atlas_tautau}
 [ATLAS Collaboration], ATLAS-CONF-2014-061.
 
     \bibitem{cms_tautau}
 [CMS Collaboration], CMS-HIG-13-004
 
\bibitem{Aad:2014lma}
  G.~Aad {\it et al.}  [ATLAS Collaboration],
 ATLAS-CONF-2014-043, arXiv:1409.3122 [hep-ex].
  
    \bibitem{atlas_ww}
  [ATLAS Collaboration], ATLAS-CONF-2014-060.
  
\bibitem{Chatrchyan:2013iaa} 
  S.~Chatrchyan {\it et al.}  [CMS Collaboration],
  JHEP {\bf 1401}, 096 (2014)
  [arXiv:1312.1129 [hep-ex]].
  
  
  \bibitem{atlas_zz}
    [ATLAS Collaboration], ATLAS-CONF-2014-044
    
\bibitem{Chatrchyan:2013mxa} 
  S.~Chatrchyan {\it et al.}  [CMS Collaboration],
  Phys.\ Rev.\ D {\bf 89}, 092007 (2014)
  [arXiv:1312.5353 [hep-ex]].
  
\bibitem{Dolan:2012rv} 
  M.~J.~Dolan, C.~Englert and M.~Spannowsky,
  JHEP {\bf 1210}, 112 (2012)
  [arXiv:1206.5001 [hep-ph]].
  
\bibitem{Baglio:2012np} 
   J.~Baglio, A.~Djouadi, R.~Gr�ber, M.~M.~M�hlleitner, J.~Quevillon and M.~Spira,
  JHEP {\bf 1304}, 151 (2013)
  [arXiv:1212.5581 [hep-ph]].
   
\bibitem{Chatrchyan:2012jja} 
  S.~Chatrchyan {\it et al.}  [CMS Collaboration],
  Phys.\ Rev.\ Lett.\  {\bf 110}, 081803 (2013)
  [arXiv:1212.6639 [hep-ex]].
   
\bibitem{Freitas:2012kw}
  A.~Freitas and P.~Schwaller,
  Phys.\ Rev.\ D {\bf 87}, no. 5, 055014 (2013)
  [arXiv:1211.1980 [hep-ph]].
  
\bibitem{Bhattacharyya:2012tj} 
  G.~Bhattacharyya, D.~Das and P.~B.~Pal,
  Phys.\ Rev.\ D {\bf 87}, 011702 (2013)
  [arXiv:1212.4651 [hep-ph]].
  
  
\bibitem{Cheung:2013kla}
  K.~Cheung, J.~S.~Lee and P.~Y.~Tseng,
  JHEP {\bf 1305}, 134 (2013);
  Phys.\ Rev.\ D {\bf 90}, 095009 (2014).
  
\bibitem{Shu:2013uua} 
  J.~Shu and Y.~Zhang,
  Phys.\ Rev.\ Lett.\  {\bf 111}, no. 9, 091801 (2013)
  [arXiv:1304.0773 [hep-ph]].
  
\bibitem{Inoue:2014nva} 
  S.~Inoue, M.~J.~Ramsey-Musolf and Y.~Zhang,
  Phys.\ Rev.\ D {\bf 89}, 115023 (2014)
  [arXiv:1403.4257 [hep-ph]].
  

\bibitem{Bolognesi:2012mm} 
  S.~Bolognesi, Y.~Gao, A.~V.~Gritsan, K.~Melnikov, M.~Schulze, N.~V.~Tran and A.~Whitbeck,
  Phys.\ Rev.\ D {\bf 86}, 095031 (2012)
  [arXiv:1208.4018 [hep-ph]].
  
\bibitem{Englert:2012xt} 
  C.~Englert, D.~Goncalves-Netto, K.~Mawatari and T.~Plehn,
  JHEP {\bf 1301}, 148 (2013)
  [arXiv:1212.0843 [hep-ph]].
  
  
\bibitem{Djouadi:2013qya} 
  A.~Djouadi and G.~Moreau,
  Eur.\ Phys.\ J.\ C {\bf 73}, no. 9, 2512 (2013)
  [arXiv:1303.6591 [hep-ph]].

 
\bibitem{Ellis:2013yxa}
  J.~Ellis, D.~S.~Hwang, K.~Sakurai and M.~Takeuchi,
  JHEP {\bf 1404}, 004 (2014)
  [arXiv:1312.5736 [hep-ph]].
  
\bibitem{Kobakhidze:2014gqa}
  A.~Kobakhidze, L.~Wu and J.~Yue,
  JHEP {\bf 1410}, 100 (2014)
  [arXiv:1406.1961 [hep-ph]].


\bibitem{Gunion:1996vv} 
  J.~F.~Gunion, B.~Grzadkowski and X.~G.~He,
  Phys.\ Rev.\ Lett.\  {\bf 77}, 5172 (1996)
  [hep-ph/9605326].

\bibitem{Bhupal Dev:2007is} 
  P.~S.~Bhupal Dev, A.~Djouadi, R.~M.~Godbole, M.~M.~Muhlleitner and S.~D.~Rindani,
  Phys.\ Rev.\ Lett.\  {\bf 100}, 051801 (2008)
  [arXiv:0707.2878 [hep-ph]].
  
\bibitem{Berge:2008wi} 
  S.~Berge, W.~Bernreuther and J.~Ziethe,
  Phys.\ Rev.\ Lett.\  {\bf 100}, 171605 (2008)
  [arXiv:0801.2297 [hep-ph]].
  
\bibitem{He:2013tia} 
  X.~G.~He, Y.~Tang and G.~Valencia,
  Phys.\ Rev.\ D {\bf 88}, 033005 (2013)
  [arXiv:1305.5420 [hep-ph]].

\bibitem{Pilaftsis:1999qt} 
  A.~Pilaftsis and C.~E.~M.~Wagner,
  Nucl.\ Phys.\ B {\bf 553}, 3 (1999)
  [hep-ph/9902371].
  
\bibitem{He:2011ws}
  X.~G.~He, G.~Valencia and H.~Yokoya,
  JHEP {\bf 1112}, 030 (2011)
  [arXiv:1110.2588 [hep-ph]].
\bibitem{Lee:2003nta}
  J.~S.~Lee, A.~Pilaftsis, M.~S.~Carena, S.~Y.~Choi, M.~Drees, J.~R.~Ellis and C.~E.~M.~Wagner,
  Comput.\ Phys.\ Commun.\  {\bf 156}, 283 (2004)
  [hep-ph/0307377].


\bibitem{Gunion:1989we}
  J.~F.~Gunion, H.~E.~Haber, G.~L.~Kane and S.~Dawson,
  Front.\ Phys.\  {\bf 80}, 1 (2000).

\bibitem{Spira:1995rr}
  M.~Spira, A.~Djouadi, D.~Graudenz and P.~M.~Zerwas,
  Nucl.\ Phys.\ B {\bf 453}, 17 (1995)
  [hep-ph/9504378].
  
    \bibitem{atlas_hbb}
 [ATLAS Collaboration], ATLAS-CONF-2014-011.
 
\bibitem{Alwall:2014hca} 
  J.~Alwall, R.~Frederix, S.~Frixione, V.~Hirschi, F.~Maltoni, O.~Mattelaer, H.-S.~Shao and T.~Stelzer {\it et al.},
  JHEP {\bf 1407}, 079 (2014)
  [arXiv:1405.0301 [hep-ph]].
  
  
\bibitem{Dittmaier:2011ti}
  S.~Dittmaier {\it et al.}  [LHC Higgs Cross Section Working Group Collaboration],
  arXiv:1101.0593 [hep-ph].

\bibitem{Beenakker:2001rj}
  W.~Beenakker, S.~Dittmaier, M.~Kramer, B.~Plumper, M.~Spira and P.~M.~Zerwas,
  Phys.\ Rev.\ Lett.\  {\bf 87}, 201805 (2001)
  [hep-ph/0107081];
  Nucl.\ Phys.\ B {\bf 653}, 151 (2003)
  [hep-ph/0211352].
  
\bibitem{Dawson:2002tg}
  S.~Dawson, L.~H.~Orr, L.~Reina and D.~Wackeroth,
  Phys.\ Rev.\ D {\bf 67}, 071503 (2003)
  [hep-ph/0211438].


\bibitem{Frixione:2014qaa} 
  S.~Frixione, V.~Hirschi, D.~Pagani, H.~S.~Shao and M.~Zaro,
  JHEP {\bf 1409}, 065 (2014)
  [arXiv:1407.0823 [hep-ph]].
  
\bibitem{Demartin:2014fia} 
  F.~Demartin, F.~Maltoni, K.~Mawatari, B.~Page and M.~Zaro,
  Eur.\ Phys.\ J.\ C {\bf 74}, no. 9, 3065 (2014)
  [arXiv:1407.5089 [hep-ph]].
  

  
\bibitem{Gunion:1996xu}
  J.~F.~Gunion and X.~G.~He,
  Phys.\ Rev.\ Lett.\  {\bf 76}, 4468 (1996)
  [hep-ph/9602226].
   

  

  
\end{thebibliography}
\end{document}